\documentclass{article}

\usepackage[preprint]{neurips_2023}

\usepackage[utf8]{inputenc} % allow utf-8 input
\usepackage[T1]{fontenc}    % use 8-bit T1 fonts
\usepackage{hyperref}       % hyperlinks
\usepackage{url}            % simple URL typesetting
\usepackage{booktabs}       % professional-quality tables
\usepackage{amsfonts}       % blackboard math symbols
\usepackage{nicefrac}       % compact symbols for 1/2, etc.
\usepackage{microtype}      % microtypography
\usepackage{xcolor}         % colors
\usepackage{graphicx}
\usepackage{microtype}
\usepackage{graphicx}
\usepackage{subfigure}
\usepackage{booktabs} % for professional tables
\usepackage{wrapfig}
\usepackage{float}
\usepackage{array}
\usepackage{amssymb}
\usepackage{pgfplots}
\pgfplotsset{compat=1.8}
\usepgfplotslibrary{statistics}
\graphicspath{  {./images/}  }
\usepackage{multirow}

\title{Sequential Transfer Learning to Decode Heard and Imagined Timbre from fMRI Data}

\author{%
  Sean Paulsen \\
  Department of Computer Science \\
  Dartmouth College \\
  Hanover, NH 03755  \\
  \texttt{paulsen.sean@gmail.com} \\
  \And
  Michael Casey \\
  Departments of CS and Music\\
  Dartmouth College \\
  Hanover, NH 03755 \\
  \texttt{mcasey@dartmouth.edu}
}

\begin{document}
%\nolinenumbers

\maketitle

\begin{abstract}
We present a sequential transfer learning framework for transformers on functional Magnetic Resonance Imaging (fMRI) data and demonstrate its significant benefits for decoding musical timbre. In the first of two phases, we pre-train our stacked-encoder transformer architecture on Next Thought Prediction, a self-supervised task of predicting whether or not one sequence of fMRI data follows another. This phase imparts a general understanding of the temporal and spatial dynamics of neural activity, and can be applied to any fMRI dataset. In the second phase, we fine-tune the pre-trained models and train additional fresh models on the supervised task of predicting whether or not two sequences of fMRI data were recorded while listening to the same musical timbre. The fine-tuned models achieve significantly higher accuracy with shorter training times than the fresh models, demonstrating the efficacy of our framework for facilitating transfer learning on fMRI data. Additionally, our fine-tuning task achieves a level of classification granularity beyond standard methods. This work contributes to the growing literature on transformer architectures for sequential transfer learning on fMRI data, and provides evidence that our framework is an improvement over current methods for decoding timbre. 
\end{abstract}

\section{Introduction}

\begin{figure}[t]

    \centering
    \includegraphics[width=0.6\textwidth]{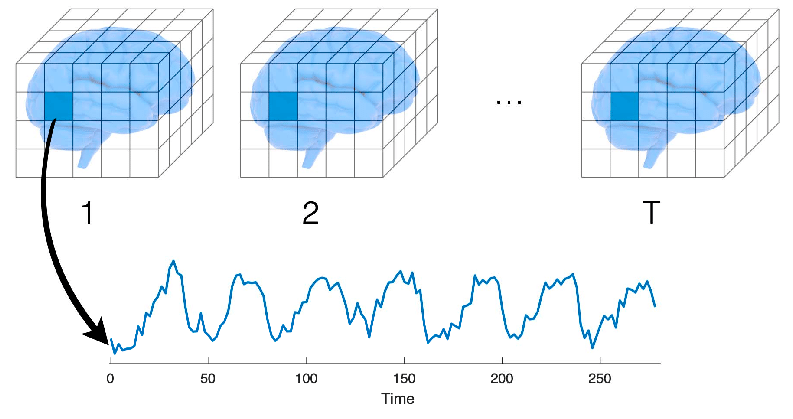}
    \caption{The sequences of voxel data used in our experiments are timeseries of neural activity measured by fMRI. Graphic published in \citet{voxelfig}}
    \label{fig:voxels}
\end{figure}

Functional MRI (fMRI) scans measure blood-oxygen-level-dependent (BOLD) responses that reflect changes in metabolic demand consequent to neural activity (\cite{bedel, hillman2014, rajapakse1998}). By measuring BOLD responses at specific combinations of spatio-temporal resolutions and coverages, fMRI data provide the means to study complex cognitive processes in the human brain (\cite{kubicki2003, wang2005, papma2017}). In particular, \textit{task-based} fMRI protocols include targeted stimuli or other task variables, such as question answering, during the scan. Researchers can then conclude associations between task features and the evoked responses across the brain (\cite{li2009, venkataraman2009, nishimoto2011}). Regions of activity that are correlated with the presence of a particular task feature are thus taken to be involved in the brain's representation of that feature (\cite{simon2004}), and they are considered to be functionally connected (\cite{rogers2007}). The discrete volumes of brain where BOLD values are measured are called \textbf{voxels} (Figure~\ref{fig:voxels}). Even rest-state fMRI data, that is, data collected in the absence of external stimuli or task, contain characteristic multi-variate signals of the brain (\cite{audimg, niu2021, yeo2011, vandijk2010, hu2006interregional}). Such rest-state signals have been shown to be predictive of the diagnosis and characterization of multiple neurological diseases and psychiatric conditions (\cite{zhan2015, woodward2015, xia2018}).

fMRI researchers have thus adopted machine learning (ML) techniques to analyze the complex relationship between BOLD signal and the underlying task, stimulus, disease, or biological information. More specifically, training an ML model to predict such information given the BOLD data as input is known as \textbf{task-state decoding}, or \textbf{brain decoding}. Toward the goal of more powerful brain decoding models, many advances in modern \textit{deep} machine learning have been applied to fMRI research. These include convolution-based models (\cite{zou2017, kawahara2017, audimg}), recurrent neural networks (RNN) (\cite{dakka2017}), and graph neural networks (\cite{li2021gnn}). Most recently, transformer (\cite{vaswani2017}) based models have achieved state of the art results on several brain decoding tasks (\cite{malkiel, bedel, nguyen}), having already grown to dominate the fields of time series forecasting (\cite{li2020timeseries}), natural language processing (\cite{devlin2019}), and computer vision (\cite{dosovitskiy2021, li2019visual}).

However, training deep models is data intensive, while fMRI scans are expensive with relatively little data obtained per scan. One strategy to somewhat alleviate the burden of data is to \textbf{pre-train} the model on a \textbf{self-supervised} task to acquire general knowledge inherent in the dataset. The pre-trained model then has a head start, so to speak, on the task of interest, by leveraging its general understanding of the data (\cite{Erhan}). This strategy is nearly ubiquitous in the domain of Natural Language Processing (NLP) (\cite{kalyan2021}) and has begun to appear in fMRI studies (\cite{nguyen,malkiel}). As \citet{kalyan2021} note, ``These models provide good background knowledge to downstream tasks which avoids training of downstream models from scratch." This process is called \textbf{sequential transfer learning}. In this paper we introduce a sequential transfer learning framework for transformers (\cite{vaswani2017}) on sequences of audio-evoked fMRI data. We then demonstrate our transformer architecture's ability to learn a novel self-supervised pre-training task and ``transfer" that knowledge to significantly improve performance on a supervised auditory brain decoding task. We further demonstrate a positive relationship between pre-training performance and fine-tuning performance, which provides further evidence that our transfer learning is genuine. These are, to the best of our knowledge, the first models to successfully decode explicit instrumental timbre. 

Our contributions are: (\textbf{1}) we present a novel self-supervised task for pre-training on sequences of fMRI data which can be applied to any dataset, (\textbf{2}) we report our transformer architecture's successful learning of this task and demonstrate transfer learning to a supervised brain decoding task, establishing a proof of concept of our pre-training task`s suitability and our framework`s capacity for transfer learning on fMRI data, and (\textbf{3}) demonstrate explicit decoding of instrumental timbre, which to our knowledge had not been done. 

\section{Related Work}
\label{gen_inst}

Univariate approaches to fMRI data such as contrast subtraction can be useful for basic analysis, but such approaches struggle to isolate the densely overlapping patterns of multivariate signals which comprise neural activity (\cite{penny2011, woolrich2001}). This challenge motivated the adoption of early ML architectures for multivariate fMRI analysis (\cite{norman2006, haxby2012}), notably support vector machines for brain decoding classification (\cite{imagined, song2014, wang2007, hojjati2017}). Progression into \textit{deep} ML models saw multilayer perceptrons (\cite{suk2016}), autoencoders (\cite{audimg, huang2018}), convolutional neural networks (CNN) (\cite{wang2020, yamins2014}), and graph neural networks (GNN) (\cite{li2021gnn}) for feature extraction and classification of single fMRI images. Time series analysis is perhaps more desirable due to the high degree of temporal correlation in BOLD responses, and indeed recurrent neural networks (RNN) and various long short-term memory (LSTM) models have been reported (\cite{dvornek2017, li2020lstm, zhao2020, thomas2019, pominova2018}).

Most recently, the transformer architecture has emerged as a superior alternative to recurrent methods for fMRI timeseries modeling. \citet{bedel} improved the state of the art for timeseries classification on multiple public fMRI datasets with a novel fused-window attention mechanism, but their work did not explore pre-training or transfer learning.  \citet{nguyen} achieved state of the art classification accuracy for a task-state decoding task on the Human Connectome Project 7-task dataset (\cite{hcp}). Their analysis includes the explicit benefits of the transformer's self-attention module when compared to previous recurrent architectures, as well as a demonstration of transfer learning when pre-training on held-out subsets of HCP 7-task. However, their pre-training task was supervised classification specific to HCP 7-task labelled data, and thus their pre-trained models would be of little to no value toward transfer learning on different datasets or modalities (\cite{kalyan2021}). \citet{malkiel} pre-trained on a self-supervised fMRI reconstruction task by wrapping the transformer block in an encoder-decoder. They report that their pre-training was crucial for improved state of the art performance on a variety of fMRI tasks such as age and gender prediction, and schizophrenia recognition. However, we note that their downstream task uses the CLS token decoding method popularized by \citet{devlin2019}, while their pre-training task does \textit{not} incorporate the CLS token. This inconsistency between training phases does not obtain the full value of the transfer learning paradigm.

Extending the above work, we explore a sequential transfer learning framework with a novel self-supervised pre-training task which includes the CLS token, can be applied to any fMRI dataset, and has all model inputs in a standardized geographical brain space \textit{without} passing through an embedding layer.

\begin{figure*}[t]

    \centering
    \includegraphics[width=1.0\textwidth]{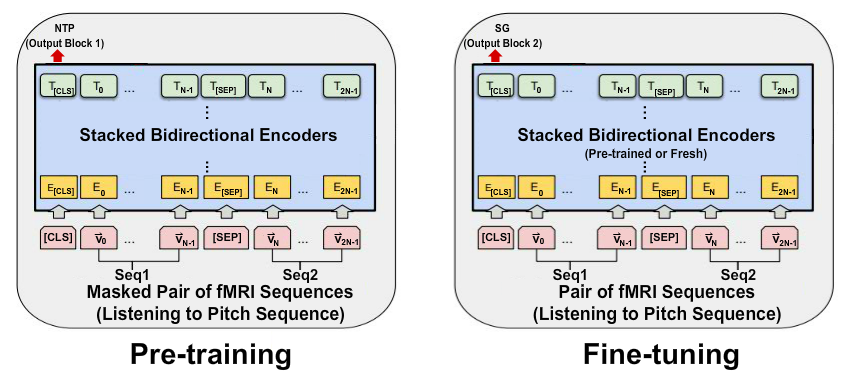}
    \caption{Pre-training and fine-tuning phases. Output Blocks are not pictured but are detailed in corresponding sections. The model learns to extract information into the CLS token, which is fed to Output Block 1 during pre-training, and Output Block 2 during fine-tuning, for classification. The SEP token separates the two sequences. For fine-tuning, the model either loads the pre-trained weights or trains a fresh model. In either case, all parameters are trained.}
    \label{fig:architecture}
\end{figure*}

\section{Architecture and Training Tasks}
\label{archandtasks}

\subsection{Paired-Sequence Transformer}
Our architecture is a modified stacked bi-directional encoder, depicted in Figure~\ref{fig:architecture}. It has two separate output blocks, one for pre-training and one for fine-tuning. The output blocks are not pictured but are detailed in their corresponding sections. We implemented our models from scratch with the pyTorch library. Our model does not include the standard embedding layer after positional encoding. We hypothesized that the composition of the fMRI scanner's measurement of BOLD signal with the preprocessing pipeline constitutes a meaningful embedding of the physical, biological neural response. The data are \textit{already} in a shared, distributed, representative space after preprocessing. Hence, we dispense with the embedding layer in our design.

\begin{figure*}[t]

    \centering
    \includegraphics[width=1.0\textwidth]{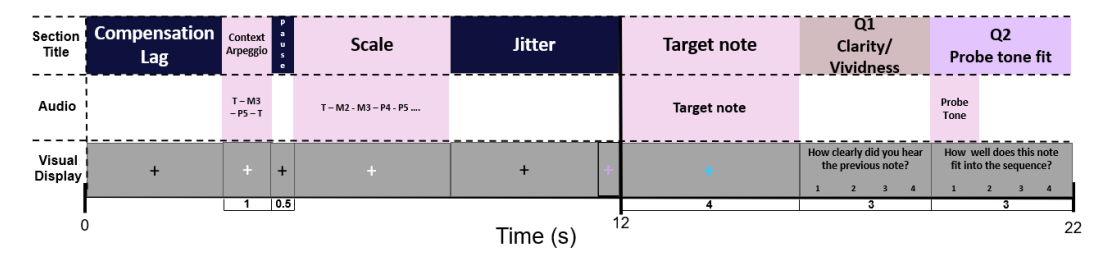}
    \caption{A single trial during scanning for the Auditory Imagery dataset, as depicted in \citet{imagined}. 21 cycles are collected from each of 8 runs for 17 participants. The arpeggio provides a tonal context, followed by a variable-length sequence of ascending notes in a major scale, then the Target Note onset indicates that either the next note in the scale is played for the participant, or the participant is prompted to \textit{imagine} the next note in the scale. The participant then rates the vividness of the Target Note whether Heard or Imagined. Finally a probe tone is played and the participant is asked how well the probe tone fits into the tonal context.}
    \label{fig:audimgprotocol}
\end{figure*}

All training data in this work were built from the Auditory Imagery fMRI Dataset (\cite{imagined}). We refer to that work for the full details of the data collection, but we summarize here. The images in this dataset were collected from seventeen participants across 8 runs in a single scan. Each run consists of 21 trials. A single trial is depicted in Figure~\ref{fig:audimgprotocol}. Each trial consists of an arpeggio to establish a tonal context, followed by a variable-length sequence of ascending notes in a major scale. At the onset of \textbf{Target Note} in the diagram, the participant either hears the next note in the scale or is prompted to imagine the next note in the scale. Next a probe tone is played, and the participant is asked to rate that tone's goodness of fit in the established tonal context. Half of each participant's runs have the Heard Target Note, the other half Imagined. Within each of those cases, half of the runs have all audio played by Clarinet, the other half by Trumpet. Therefore 2 runs each are designated as Heard Clarinet, Heard Trumpet, Imagined Clarinet, and Imagined Trumpet. We refer to these designations as the four ``conditions," while Trumpet and Clarinet are the two ``instrumental timbres".

Before constructing the inputs, we performed the fmriprep (\cite{fmriprep}) preprocessing pipeline on the source dataset via the Brainlife website (\cite{brainlife}). The final step of this pipeline is to warp all of the data into a standard geographical space so that different brains can be directly compared. We then standardized the data to mean 0 and standard deviation 1, followed by linear de-trending across all runs. Finally we reduced all fMRI images to only the left-side auditory cortex using the Harvard-Oxford cortical atlas, resulting in 420 voxels for each image, which we then flattened to 1-D. Left side was chosen arbitrarily, thus right and bilateral auditory cortex remain for future work. Each input to the model is constructed by extracting a contiguous sequence of fMRI images from a participant's data (\textbf{Seq1}), and pairing it with another such sequence (\textbf{Seq2}), as seen in Figure~\ref{fig:architecture}. As a starting point for this line of work we wanted to keep training times lower and extract as much training data as possible by having relatively short sequences. Thus the length of each of the two sequences in an input is 5. We refer to such a sequence as a \textbf{5-seq} throughout this work. 

A \textbf{separator token} (SEP) (\cite{devlin2019}) is inserted between the two sequences, and a \textbf{classification token} (CLS) (\cite{devlin2019, malkiel, nguyen}) is inserted at the front to serve as a pooling token which extracts information from the rest of the sequence. The distributed representations of these tokens are commonly learned by an embedding layer, but recall our hypothesis that the preprocessing of the fMRI data serves in place of an embedding for fMRI data. Therefore we must manually determine the 420-dimensional representation of CLS and SEP. Intuitively these tokens ought to be ``far away" from the rest of the data in the distributed space. Thus we simply reserved the first two of the 420 dimensions for our tokens. The CLS and SEP tokens have a 1 in the zeroth and first dimensions respectively, and are zero elsewhere. In order for the fMRI images to have zeros in these dimensions, we removed the two voxels with lowest probability of inclusion in left auditory cortex according to the Harvard-Oxford atlas. Thus in practice we had 418 voxels in each image rather than 420, with the two leftover dimensions reserved for the tokens. We chose this rather than increasing to 422 because the number of attention heads in the transformer layers must evenly divide the dimension of the input. The diverse factorization of 420 allowed us to test many different numbers of attention heads, while 422 would not have.

\subsection{Pre-training Tasks} 
We now present our novel self-supervised pre-training task, Next Thought Prediction (\textbf{NTP}). The goal of this task is binary classification, predicting whether or not Seq2 follows immediately after Seq1 in the original data. From the output of the final transformer block, the transformed CLS token is sent to \textbf{Output Block 1}. This block consists of a linear layer projecting down from 420 dimensions to 210, then a linear layer projecting down from 210 to 2, and finally a softmax is applied to obtain probabilities for ``No" (index 0) and ``Yes" (index 1). The loss for NTP is calculated as the Cross-Entropy between the result of Output Block 1 and a one-hot encoding of the ground truth. We hypothesized that this task would teach the model a general understanding of the temporal and spatial dynamics of neural activity, which could then be transferred to the downstream task. 

The training data for this task is created by selecting every possible 5-seq (with stride of 2) from all participants as Seq1, then for each of those, Seq2 is taken from the same participant as Seq1. For the NTP task, Seq2 either immediately follows Seq1 in the participant's data (a positive sample), or not (a negative sample). In the negative case, Seq2 is chosen uniformly at random from all 5-seqs from that participant which do not overlap with Seq1. Observe that we must choose whether to create \textit{both} a positive and negative sample for each Seq1, which could contribute to overfitting via repeated exposure, or only one of the two with 50/50 chance, which could contribute to underfitting by halving the total training data compared to the former case. We call these cases \textbf{True PosNeg} and \textbf{False PosNeg} respectively, where True and False indicate whether we created both a positive and negative sample or not. Both cases are considered in our experiments. 

\subsection{Fine-tuning Task}
Our novel supervised brain decoding task used for fine-tuning is the Same Timbre (\textbf{ST}) task. The goal of this task is binary classification, predicting whether or not Seq1 corresponds to listening to the same instrumental timbre (Clarinet or Trumpet) as Seq2. From the output of the final transformer block, the transformed CLS token is sent to Output Block 2. This block consists of a single linear layer projecting down from 420 dimensions to 2, then a softmax is applied to obtain probabilities for ``No" (index 0) and ``Yes" (index 1). The loss for ST is calculated as the Cross-Entropy between the result of Output Block 2 and a one-hot encoding of the ground truth. We made the fine-tuning output block as simple as possible to ensure that the brunt of the work is supported by the pre-trained transformer blocks.

The training data for this task must match the architecture used in pretraining, so once again the inputs are pairs of 5-seqs. For this task, the 5-seqs are the five contiguous images beginning from the Target Note onset from each cycle (Figure~\ref{fig:audimgprotocol}). We wanted to include the Imagined Target Notes, but collecting 5-seqs from the arpeggio or the scale would bias the model toward Heard sound, so we did not collect those. Because there are drastically fewer desired 5-seqs than we had in NTP, we prioritized the size of the training set and did not consider False PosNeg. In other words, every 5-seq was used as Seq1 to construct both a positive and negative sample. As in NTP, Seq2 was chosen within-participant. In all pairings, Seq1 and Seq2 are either both Heard or both Imagined. We leave the Heard-Imagined cross pairs to future work.

\section{Experiments and Results}
\label{experiments}

\subsection{Pre-training} 
Recall that the source dataset consists of two each of Heard Clarinet, Heard Trumpet, Imagined Clarinet, and Imagined Trumpet. Holding out any single run as a validation set would not express the model's ability to learn all four conditions. If we hold out a strict subset from each of the four conditions, then some runs will contribute to both training and validation splits, raising contamination concerns. Therefore we (reluctantly) held out half the total data, that is, the collection of one run of each condition from each participant, as a validation split. Thus we obtained 7548 training and validation samples for False PosNeg, and 15096 training and validation samples for True PosNeg.

We performed a basic hyperparameter grid search over the Learning Rate, the number of Transformer Layers, the number of Attention Heads within each layer, and the factor of Forward Expansion within each layer. The best performing configuration on the held out data was, in that order, [$10^{-5}$, $3$, $2$, $4$].

With the above hyperparameters, we trained 10 models on the False PosNeg data and 10 models on the True PosNeg data. We applied a dropout rate of 0.1 in all transformer layers. Models were trained for ten epochs via backpropagation with the Adam optimizer with $\beta_1=0.9, \beta_2=0.999,$ and weight\_decay $=0.0001$. Each model was trained with a different RNG seed for reproducibility. 

Results are presented in Table~\ref{table:pre-trainaudimg}. For each training session, we saved the model's state after the epoch with the highest NTP accuracy on the validation split (``Best Val Acc" in the table). The ``Best Epoch" column contains the epoch in which this accuracy was achieved, from 0 to 9. The averages of each column are given in the last row of the table.

\begin{table}[t]
    \begin{center}
    \caption{Results of pre-training 10 models on NTP with the Auditory Imagery Dataset. One run of each condition was held out for validation. True PosNeg means that each 5-seq was used as Seq1 for both a positive and negative sample. False PosNeg means either positive or negative with 50/50 chance. Best Val Acc is the highest accuracy obtained during training on the validation split. The epoch in which that accuracy was obtained is given in the Best Epoch column, from 0 to 9 inclusive. The average across all ten models is given at the bottom of each column.}
    \label{table:pre-trainaudimg}
    \begin{tabular}{|| m{1.5cm}| m{2.25cm} | m{1.25cm} | m{2.2cm} | m{1.2cm} ||}
    \hline
    & \multicolumn{2}{|c|}{\textbf{True PosNeg}}& \multicolumn{2}{|c|}{\textbf{False PosNeg}}\\
    \hline
    \textbf{\footnotesize{Iteration}} & \textbf{Best Val Acc} & \textbf{\footnotesize{est Epoch}} & \textbf{\footnotesize{Best Val Acc}}   & \textbf{Best Epoch}  \\
    \hline
    0 & 86.116\% & 8 & 74.232\% & 9    \\

    \hline
    1 & 88.003\% & 9 & 78.259\% & 5   \\

    \hline
    2 & 83.797\% & 8 & 76.683\% & 9  \\

    \hline
    3 & 90.196\% & 9 & 85.533\% & 9   \\

    \hline
    4 & 88.408\% & 7 & 82.764\% & 9  \\

    \hline
    5 & 85.102\% & 6 & 87.228\% & 9   \\

    \hline
    6 & 87.056\% & 9 & 76.312\% & 4   \\

    \hline
    7 & 88.116\% & 8 & 74.271\% & 8   \\

    \hline
    8 & 85.182\% & 9 & 81.373\% & 9   \\

    \hline
    9 & 89.388\% & 9 & 79.518\% & 9  \\

    \hline
     Average & 87.136\% & 8.2 & 79.609\% & 8.0 \\

    \hline

    \end{tabular}

    \end{center}
\end{table}

All models performed well above chance. Recall our concern about potential overfitting due to repeat exposure in the True PosNeg case. Our results are rather to the contrary. True PosNeg obtains significantly higher performance over False PosNeg ($p=0.0002$), although the amount of training epochs required to reach that performance is almost identical to False PosNeg.

\subsection{Fine-tuning}
For fine-tuning, we constructed the training data as described previously with the same runs held out for validation as in pre-training to avoid contamination. Thus we obtained 2856 training and validation samples. We fine-tuned ten models by loading the ten saved Best Epoch weights from the True PosNeg case, another ten after loading the saved weights from False PosNeg. Preliminary testing showed that freezing the pre-trained weights and updating only Output Block 2 was not a viable training regimen. Therefore all parameters were updated during fine-tuning. To examine the benefit of transfer learning, we also trained ten ``fresh" models. The fresh models are identical to the other models used in fine-tuning but do not load any pre-trained weights. We also trained ten null models, which are identical to fresh models but have the True/False labels assigned uniformly at random to the training data. In the interest of brevity, we simply report here that the average best performance of the null models was roughly $51\%$ rather than reporting their individual results.

Models were trained for ten epochs via backpropagation with the same Adam parameters and hyperparameters as during pre-training. Each model was trained with a different RNG seed for reproducibility. Table~\ref{table:fine-tuneaudimg1} contains the results. We remind the reader that the PosNeg columns in this table refer to which pre-trained weights were loaded, \textit{not} the construction of the dataset in this phase.

\begin{table*}[t]
    \begin{center}
    \caption{Results of fine-tuning after loading the Best Epoch weights from the ten True/False PosNeg models, as well as ten fresh models which serve as a baseline to examine the effects of transfer learning. The average across all ten models is given at the bottom of each column.}
    \label{table:fine-tuneaudimg1}
    \begin{tabular}{||m{1.5cm}| m{1.8cm} | m{1.2cm} | m{1.8cm} | m{1.2cm} | m{1.8cm} | m{1.2cm} ||}
    \hline
    & \multicolumn{2}{|c|}{\textbf{True PosNeg}}& \multicolumn{2}{|c|}{\textbf{False PosNeg}} & \multicolumn{2}{|c||}{\textbf{Fresh}}\\
    \hline
    \textbf{\footnotesize{Iteration}} & \textbf{\footnotesize{Best Val Acc}} & \textbf{Best Epoch} & \textbf{\footnotesize{Best Val Acc}}   & \textbf{Best Epoch}  & \textbf{\footnotesize{Best Val Acc}} & \textbf{Best Epoch}\\
    \hline
    0 & 70.553\% & 4 & 68.277\% & 4  &  65.966\%   &  4   \\

    \hline
    1 & 73.915\% & 8 & 71.849\% & 2  &  67.752\%   &  5   \\

    \hline
    2 & 75.210\% & 6 & 70.868\% & 8 &  66.527\%    &  7   \\

    \hline
    3 & 77.906\% & 3 & 73.389\% & 4  &  66.632\%    &  7   \\

    \hline
    4 & 76.120\% & 6 & 71.148\% & 3 &  67.017\%   &  6   \\

    \hline
    5 & 76.366\% & 6 & 76.331\% & 9  &  68.768\%   &  9  \\

    \hline
    6 & 74.545\% & 0 & 71.744\% & 9  &  70.238\%    &  9   \\

    \hline
    7 & 71.254\% & 9 & 69.223\% & 5  &  67.752\%    &  8   \\

    \hline
    8 & 74.615\% & 2 & 69.538\% & 9  &  62.29\%    &  6   \\

    \hline
    9 & 75.490\% & 2 & 70.168\% & 9  &  64.321\%    &  8   \\

    \hline
     Average & 74.597\% & 4.60 & 71.254\% & 6.2 & 66.726\% & 6.9\\

    \hline

    \end{tabular}

    \end{center}
\end{table*}

The models fine-tuned after loading the best False PosNeg weights significantly outperformed the fresh models ($p=0.0003$), and the models fine-tuned on True PosNeg weights significantly outperformed False PosNeg ($p=0.0042$). Further, they appear to reach their best performance with less training than the others, but this is not conclusive.

\subsection{Discussion}
The first point of interest is that the pre-training phase was successful at all. fMRI data is a challenging domain and our work is the first to use paired-sequence transformers within that domain. Nevertheless, our models significantly outperformed the 50/50 random chance on Next Thought Prediction. \textbf{We contribute this result as significant evidence that the Next Thought Prediction task is well-defined and our novel paired-sequence architecture is capable of learning it}. 

Our novel supervised task, Same Timbre, was also successful on both pre-trained models and fresh models, while the null models performed only at random chance. \textbf{We contribute this result as significant evidence that the Same-Timbre task is well-defined and our model is suited to learn it}. Note that the success in pre-training and fine-tuning also validates our novel implementation of the CLS and SEP tokens. Further, the pre-trained models significantly outperformed the fresh models. \textbf{We contribute this result as significant evidence of the ability to perform sequential transfer learning with our framework}. Additionally, one would intuitively expect a positive relationship between pre-training performance and fine-tuning performance. Indeed, we obtained significantly higher average performance on the ST task when the models loaded the True PosNeg weights, that is, the weights with higher average performance on NTP. \textbf{We contribute this result as evidence supporting our hypothesis that Next Thought Prediction teaches the model a general understanding of the temporal and spatial dynamics of neural activity that can be leveraged toward downstream tasks.}

Observe that our implementation of the Same-Timbre task implicitly includes the task of decoding the Clarinet and Trumpet labels from a single 5-seq. For example, if we want to predict whether a Heard 5-seq is Clarinet or Trumpet, we need only pair it with a 5-seq whose Clarinet/Trumpet label is known, and then ask the model if they are the same timbre. Thus \textbf{we contribute our success on the Same-Timbre task as the first successful decoding of instrumental timbre from auditory cortex in fMRI data.}

\section{Conclusions and Future Work}

In this work we presented a novel sequential transfer learning framework for sequences of fMRI data. This framework includes our paired-sequence transformer architecture, the Next Thought Prediction self-supervised pre-training task, and the Same-Timbre task, our downstream supervised brain decoding task. Our pre-training and fine-tuning results provide significant evidence that \textbf{(1)} our tasks are well-defined, \textbf{(2)} that our architecture is suitable for learning them, and \textbf{(3)} that our framework facilitates sequential transfer learning. Our finetuning results additionally provided evidence that \textbf{(4)} our framework is able to decode instrumental timbre, which to our knowledge had not been done.

Looking forward, note the similarity between the Same Timbre task and a standard contrast analysis of timbre in STG. Fundamentally, both are asking if STG encodes different genres in different ways. Our implementation answers this question in the affirmative without the need for a contrast analysis. More generally, any contrast analysis interested in the difference between two conditions could be substituted by our model. We believe this paired-sequence framework has potential for replacing or supplementing contrast analysis, but there is still a great deal of work with fMRI data that cannot use the paired-sequence structure presented in this work. Our upcoming work has a novel self-supervised pre-training task for only \textit{one} sequence, which generalizes immediately to common brain decoding tasks and datasets.

Looking elsewhere, the experiments in this work do not exploit a particular power of transformers for learning long term dependencies. Moreover, the hemodynamic response function is known to peak 4 to 6 seconds \cite{bonakdarpour2007} (2-3 images in our case) after stimulus onset, so the peak response to the probe tone (presented before the fifth TR of each 5-seq) is not seen by the model. For both of these reasons, longer sequences are of interest in future study. 

We hypothesize that this transfer learning framework can be extended to other ROIs as well as whole brain analysis. Of particular interest would be a pre-trained model that can transfer knowledge to different datasets. This would require a pre-training dataset with general tasks across the whole brain. The Human Connectome Project (2013) \cite{hcp} is a prevailing candidate for this future work due to its overwhelming size and well-established benchmarks across the entire brain.

\nocite{*}

\bibliographystyle{plainnat}
\bibliography{ptlbib}

%%%%%%%%%%%%%%%%%%%%%%%%%%%%%%%%%%%%%%%%%%%%%%%%%%%%%%%%%%%%

\end{document}